\def\be{\begin{equation}}
\def\ee{\end{equation}}
\def\ba{\begin{eqnarray}}
\def\ea{\end{eqnarray}}
\def\no{\nonumber \\}
\newcommand{\nid}{\noindent}

\documentclass[12pt]{article}
\usepackage{bm}

\begin{document}

\begin{flushright}
NIIG-DP-06-2\\
hep-lat/0609025
\end{flushright}

\begin{center}
{\LARGE \bf
Rotational Symmetry and A Light Mode in \\
\vskip .1in
Two-Dimensional Staggered Fermions 
}\\
\vskip .75in

{\large
Morio \textsc{Hatakeyama},$^{a}$\footnote{E-mail: hatake@muse.sc.niigata-u.ac.jp}
Hideyuki \textsc{Sawanaka}$^{a}$\footnote{E-mail: hide@muse.sc.niigata-u.ac.jp}
and
Hiroto \textsc{So}$^{b}$\footnote{E-mail: so@muse.sc.niigata-u.ac.jp}
}\\
\vskip .2in

{\it $^a$Graduate School of Science and Technology,} \\ 
{\it Niigata University}, \\ 
{\it Ikarashi 2-8050, Niigata 950-2181, Japan}
\\
{\it $^{b}$Department of Physics},\\
{\it Niigata University},\\
{\it Ikarashi 2-8050,
 Niigata 950-2181, Japan.}\\

\vskip .5in
\end{center}

\baselineskip 24pt
\thispagestyle{empty}

\begin{abstract}
\baselineskip 16pt
To obtain a light mode in two-dimensional staggered fermions, 
we introduce four new local operators keeping 
the rotational invariance for a staggered Dirac operator. To split 
masses of tastes, three cases  are considered.  
The mass matrix and the propagator for  free theories are analyzed.  
We find that one of three cases is a good candidate for 
taking a single mode by the mass splitting.  
In the case, it is possible that a heavy mode obtains infinite mass 
on even sites or odd sites. 
\end{abstract}


\setcounter{footnote}{0}

\newpage
\section{Introduction}    
Staggered fermions are formulated 
in which  species doublers of a Dirac field 
are interpreted as physical degrees of freedom, {\it tastes}, on 
lattice~\cite{KS,Susskind}.   Although  the fermion determinant 
has  many advantages in the cost of 
its numerical calculation~\cite{HPQCD,MILC}, 
it remains for a 4-fold degeneracy problem  of tastes in four dimensions 
to be unsolved.  
A fourth-root trick of the determinant 
in a staggered Dirac operator  is  an approach to unfold the degeneracy 
and studies on  its theoretical basis are developed~\cite{Adams:2003rm,Shamir,Giedt}.  
However,  we have no local expression of  one taste Dirac fermion 
after the fourth-root trick. 

Avoiding the trick, there are pioneering works 
for solving the degeneracy tried 
by improved staggered fermion approaches~\cite{Mitra,Hart}.  
The improved actions generally include more operators than 
the original staggered one and are difficult to treat them~\cite{Toolkit}.  
For the control of their operators, 
we make use of staggered fermions 
on a $D$-dimensional lattice space based on 
an $SO(2D)$ Clifford algebra~\cite{IKMSS}.  The formulation 
by the $SO(2D)$ Clifford algebra is powerful in the control   
of  fermion operators  and {\it we can describe 
any improved  fermion action on a hypercubic lattice}.  In addition, 
a discrete rotational symmetry (cubic symmetry) can be represented by the algebra.  

In this article, we analyze the mass splitting of degenerate tastes 
by adding four operators to the original staggered action in two dimensions.  
Only these four operators keep the discrete rotational symmetry 
in any dimension~\cite{IKMSS}.  The total mass matrix analysis 
is insufficient because  the matrix  does not commute with  
the kinetic term.  Therefore, we also analyze 
the propagator and the pole.  It is found that only one combination 
in these operators is a good candidate after these analyses. 

This article is organized as follows.  In section 2,  
staggered fermions are formulated 
by the $SO(2D)$ Clifford algebra.  Four operators are introduced 
to obtain taste-splitting masses.  These operators keep 
the discrete rotational invariance.  We analyze the mass matrix 
and the mass pole of the improved free staggered Dirac operator 
in sections 3 and 4.  Section 5 is devoted to further analyses of 
the massless limit and infinite mass of a heavy mode.  We 
summarize  and discuss about our approach in section 6.

\section{Formulation of Staggered Fermions and  Rotational Symmetry} 
\begin{figure}
\begin{center}
 \begin{picture}(400,130)
 \linethickness{0.3mm}
 \put(180,30){\line(0,1){70}} \put(250,30){\line(0,1){70}}
 \put(180,30){\line(1,0){70}} \put(180,100){\line(1,0){70}}
 \put(180,30){\circle*{5}} \put(180,100){\circle*{5}}
 \put(250,30){\circle*{5}} \put(250,100){\circle*{5}}
 \put(120,15){$\bm (-1/2,-1/2)$} \put(128,110){$\bm (-1/2,1/2)$}
 \put(250,15){$\bm (1/2,-1/2)$} \put(250,110){$\bm (1/2,1/2)$}
 \put(215,10){\vector(0,1){110}} \put(160,65){\vector(1,0){110}}
 \put(213,125){$\bm r_2$} \put(275,57){$\bm r_1$} \put(205,55){$\bm O$}
 \end{picture}
 \caption{A two-dimensional lattice unit and  the weight of  a spinor 
representation in $SO(4)$.}
 \label{unit-2dim}
\end{center}
\end{figure}
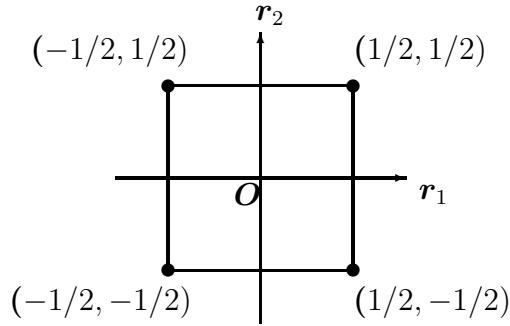
The formulation of staggered fermions 
on the $D$-dimensional lattice space has been presented 
based on the $SO(2D)$ Clifford algebra~\cite{IKMSS}.  The basic idea 
is that the dimension of the total representation space 
including  spinor and taste spaces, $2^D$ 
is the same as that of  an $SO(2D)$ spinor representation.  $2^D$ 
is also the same as the number of sites in a $D$-dimensional 
hypercube.  To avoid the double counting of sites, 
the lattice coordinate $n_{\mu}$  is  noted by 
\be
 n_{\mu} = 2 N_{\mu}+ c_{\mu} + r_{\mu},  
\ee
\nid
where $N_{\mu}$ is the global coordinate of the hypercube.  
In this case, a fundamental unit is $2a$, where $a$ is a lattice 
constant, and is set to unity.  $ c_{\mu} = 1/2$ for any $\mu$ 
means  the coordinate of a center in the $D$-dimensional hypercube and 
$r_\mu$ does the relative coordinate of a site to the center.  
The relative coordinate is the same as  a weight of  the spinor 
representation in  $SO(2D)$.  

Although our formulation can be generalized,  we consider a theory 
in a two-dimensional lattice, for simplicity.  Relative 
coordinates of four sites around a plaquette are written by 
\be
 (r_1, r_2) = (-1/2, -1/2),~ (-1/2, 1/2),~ (1/2, -1/2),~ (1/2, 1/2) ,
\ee
\label{4sites}
\nid
as shown in Figure~\ref{unit-2dim}. 
Actually, our staggered fermion is defined on sites (\ref{4sites}) as 
\be
 \Psi(n) \equiv \Psi_{r}(N) = \left(\begin{array}{c} 
  \Psi_{(-1/2,-1/2)} \\ \Psi_{(-1/2,1/2)} \\ 
  \Psi_{(1/2,-1/2)} \\ \Psi_{(1/2,1/2)} \end{array}\right) (N) 
 \equiv  \left(\begin{array}{c} 
  \Psi_1 \\ \Psi_2 \\ \Psi_3 \\ \Psi_4 \end{array}\right) (N). 
\ee
\nid
It is noted that $\Psi_1$ and  $\Psi_4$ are put on even sites and 
$\Psi_2$ and  $\Psi_3$ are put on odd sites.  

An $SO(4)$ Clifford algebra plays a crucial role in 
two-dimensional cubic lattice formulations~\cite{IKMSS}.  The original 
staggered fermion action~\cite{KS,Susskind} can be written as 
\be
 S_{st} = \sum_{N,N^\prime,r,r^\prime,\mu,\vec{\tau}} 
 \bar{\Psi}_r(N)(D_{\mu}^{\vec{\tau}})_{(N,N^\prime)} 
 (\Gamma_{\mu,\vec{\tau}})_{(r,r^\prime)}\Psi_{r^\prime}(N^\prime) , 
\label{action}
\ee
\nid
where $\vec{\tau}$ is a two-dimensional vector 
with its components of $\pm 1/2$  and 
$ D_{\mu}^{\vec{\tau}}$ for $\mu = 1, 2$, is 
a generalized difference  operator defined by 
\be
 (D_{\mu}^{\vec{\tau}})_{(N,N^\prime)} \equiv 
 \frac{1}{2^2}\sum_{\vec{\sigma}=0,1} 
 (-1)^{(\vec{c}+\vec{\tau})\cdot\vec{\sigma}} 
 (\nabla_\mu^{\vec{\sigma}})_{(N,N^\prime)} , 
\ee
\nid
with
\be
 (\nabla_\mu^{\vec{\sigma}})_{(N,N^\prime)} = \left\{\begin{array}{ll} 
  \delta_{N,N^\prime}U_{2N+\vec{\sigma},\mu} - 
  \delta_{N-\hat{\mu},N^\prime}U_{2N+\vec{\sigma}-\hat{\mu},\mu}^\dag 
  \equiv \nabla_\mu^-, & ~\sigma_\mu=0, \\
 \\ 
  \delta_{N+\hat{\mu},N^\prime}U_{2N+\vec{\sigma},\mu} - 
  \delta_{N,N^\prime}U_{2N+\vec{\sigma}-\hat{\mu},\mu}^\dag 
  \equiv\nabla_\mu^+, & ~\sigma_\mu=1. 
 \end{array}\right. 
\label{gen-diff}
\ee
\nid
$\vec{\sigma}$ is a two-dimensional vector dual to $\vec{\tau}$ and 
$\nabla_\mu^+$ ($\nabla_\mu^-$) implies 
a forward (backward) difference operator along the $\mu$-direction, 
respectively.  In Eq.~(\ref{gen-diff}), a link variable 
$U_{2N+\vec{\sigma},\mu}$ is introduced for gauge covariance.  
This formulation of staggered fermions  has usual gauge interactions 
and  definite  $O(a)$  higher order terms.  The matrix 
$\Gamma_{\mu,\vec{\tau}}$ in our action (\ref{action}) is 
composed of the $SO(4)$ Clifford algebra $\gamma_{\mu}$ and 
$\tilde{\gamma}_{\mu}$ 
\be
 \Gamma_{1, -\vec{c}} = \left(\begin{array}{cccc}
  0 & 0 & 1 & 0 \\
  0 & 0 & 0 & 1 \\
  1 & 0 & 0 & 0 \\
  0 & 1 & 0 & 0 \end{array}\right) \equiv \gamma_1, \quad 
 \Gamma_{2, -\vec{c}} = \left(\begin{array}{cccc}
  0 & 1 &  0 &  0 \\
  1 & 0 &  0 &  0 \\
  0 & 0 &  0 & -1 \\
  0 & 0 & -1 &  0 \end{array}\right) \equiv \gamma_2,
\ee
\be
 -i \Gamma_{1, (1/2,-1/2)} = \left(\begin{array}{cccc}
  0 & 0 & -i &  0 \\
  0 & 0 &  0 & -i \\
  i & 0 &  0 &  0 \\
  0 & i &  0 &  0 \end{array}\right) \equiv \tilde{\gamma}_1, \quad 
 -i \Gamma_{2, (-1/2,1/2)} = \left(\begin{array}{cccc}
  0 & -i &  0 & 0 \\
  i &  0 &  0 & 0 \\
  0 &  0 &  0 & i \\
  0 &  0 & -i & 0 \end{array}\right) \equiv \tilde{\gamma}_2,
\ee
\nid
and  is described by 
\be
 \Gamma_{\mu,\vec{\tau}} \equiv 
 (i\tilde{\gamma}_1\gamma_1)^{1/2+\tau_1} 
 (i\tilde{\gamma}_2\gamma_2)^{1/2+\tau_2} \gamma_{\mu}. 
\ee
\nid
Here we denote the fundamental algebra, or 
the $SO(4)$ Clifford algebra as 
\be
 \{\gamma_{\mu},\gamma_{\nu}\} = 
 \{\tilde{\gamma}_{\mu},\tilde{\gamma}_{\nu}\} = 2\delta_{\mu\nu},
\ee
\be
 \{\gamma_{\mu},\tilde{\gamma}_{\nu}\}=0, 
\ee
\nid
and 
\be
 \{\Gamma_{\mu,\vec{\tau}},  \Gamma_5\} = 0 , 
\ee
\nid
where 
$\Gamma_5 \equiv \gamma_1\gamma_2 \tilde{\gamma}_1\tilde{\gamma}_2 
 = diag(1,-1,-1,1)$.  {}From the algebra,  we find that 
a massless staggered fermion has an even-odd symmetry 
\be
 \Psi \rightarrow e^{i\theta \Gamma_{5}}\Psi, \quad
 \bar{\Psi} \rightarrow \bar{\Psi}e^{i\theta \Gamma_{5}}.
\ee

For a discrete rotation with angle $\pi/2$  around the center, 
the transformations of global and relative coordinates are denoted by 
\be
 N \rightarrow  R(N), \quad r \rightarrow  R(r),
\ee
\nid
and that of fermion  is 
\be
 \Psi(N) \rightarrow V_{12} \Psi(R(N)).
\ee
\nid
$V_{12}$ is  a rotation matrix about a spinor index 
in the $SO(4)$ base, up to a phase factor given by a form 
\be
 V_{12} = \frac{e^{i\vartheta}}{2} \Gamma_5 
 (\tilde{\gamma}_1-\tilde{\gamma}_2)(1+\gamma_1\gamma_2) = 
 e^{i\vartheta} \left(\begin{array}{cccc} 
  0 & 0 & -i & 0 \\ 
  i & 0 &  0 & 0 \\ 
  0 & 0 &  0 & i \\ 
  0 & i &  0 & 0 \end{array}\right). 
\ee

The following four operators $O_i$ for $i=1, 2, 3, 4$, 
\be
 O_1 = \bm{1}, \quad 
 O_2 = i\gamma_1\gamma_2 \equiv \Gamma_3, \quad 
 O_3 = \tilde{\gamma}_1 +  \tilde{\gamma}_2, \quad 
 O_4 = \Gamma_3 (\tilde{\gamma}_1 + \tilde{\gamma}_2), 
\ee
\nid
are invariant under the rotation $V_{12}O_i V_{12}^\dag$.  
Our analyses in the following sections concentrate on 
the improved staggered fermion action  by these four operators 
with $U_{2N+\vec{\sigma},\mu}=1$.

\section{Analysis of Mass Matrices} 
To split  masses  in degenerate tastes  we introduce four
rotationally invariant operators, which we denote as 
${\mathcal M}_i  \equiv \bar{\Psi} O_i \Psi$~\cite{IKMSS}, for 
the original staggered fermion action~(\ref{action}).  
Explicit expressions for ${\mathcal M}_i$ are given by 
\ba
 \mathcal M_1(N) &=& \sum_{r, r^\prime} \bar{\Psi}_r(N) 
  \bm{1}_{r, r^\prime}{\Psi}_{r^\prime} (N), \no 
 \mathcal M_2(N) &=& \sum_{r, r^\prime} \bar{\Psi}_r(N) 
  (\Gamma_{3})_{r, r^\prime}{\Psi}_{r^\prime} (N), \no 
 \mathcal M_3(N) &=& \sum_{r, r^\prime} \bar{\Psi}_r(N) 
  (\tilde{\gamma}_1 + \tilde{\gamma}_2)_{r, r^\prime} 
   {\Psi}_{r^\prime} (N), \no 
 \mathcal M_4(N) &=& \sum_{r, r^\prime} \bar{\Psi}_r(N) 
  \{\Gamma_{3} (\tilde{\gamma}_1 + \tilde{\gamma}_2) \}_{r, r^\prime} 
   {\Psi}_{r^\prime} (N). 
\label{4rot-inv-ope}
\ea
\nid
The total mass matrix form  which is invariant under the rotation 
by $\pi/2$ in two dimensions is given as 
\ba
 M_R &=& m_1 \bm{1} + m_2 \Gamma_3 
       + m_3 (\tilde{\gamma}_1 + \tilde{\gamma}_2) 
       + m_4 \Gamma_3(\tilde{\gamma}_1 + \tilde{\gamma}_2) \no 
     &=& \left(\begin{array}{cccc} 
       m_1 & -im_3+m_4 & -im_3-m_4 &     -im_2 \\ 
  im_3+m_4 &       m_1 &     -im_2 & -im_3+m_4 \\ 
  im_3-m_4 &      im_2 &       m_1 &  im_3+m_4 \\ 
      im_2 &  im_3+m_4 & -im_3+m_4 &       m_1 \end{array}\right), 
\ea
\nid
where $m_1, m_2, m_3, m_4$ are parameters of each operator 
in Eq.~($\ref{4rot-inv-ope}$).  $M_R$ has four eigenvalues  
\ba
 m_1-m_2-\sqrt{2}m_3+\sqrt{2}m_4 , & & 
 m_1-m_2+\sqrt{2}m_3-\sqrt{2}m_4 , \no 
 m_1+m_2-\sqrt{2}m_3-\sqrt{2}m_4 , & & 
 m_1+m_2+\sqrt{2}m_3+\sqrt{2}m_4 . 
\label{eigenvalues}
\ea
\nid
The operator  ${\mathcal M}_1$  cannot separate  
the 2-fold degeneracy  in two-dimensional staggered Dirac fermions.  
The degeneracy can be solved by remained three operators  
${\mathcal M}_2$, ${\mathcal M}_3$ and  ${\mathcal M}_4$.  
A  4-component spinor should be separated into two 
2-component spinors since a two-dimensional Dirac spinor is composed 
of a 2-component mode  and we keep the rotational invariance 
even under  a finite lattice constant%
\footnote{If one permits  the rotational invariance only after taking 
the continuum limit,  it is not necessary for degeneracy of a heavy
mode  and there are six more cases derived from Eq.~(\ref{eigenvalues}).}. 
Actually all possibilities of this separation are three cases 
and are listed in Table~\ref{3C}.  
Let us analyze these mass matrices in order explicitly. 
\begin{table}
\begin{center}
\renewcommand{\arraystretch}{1.25}
 \begin{tabular}{|c|c|c|c|} \hline 
   & parameter conditions & rotationally invariant mass term & 
  mass eigenvalues \\ \hline 
  case~ 1 & $m_2 = m_3 = 0$ & 
  $M_{R1}=m_1\bm{1}+m_4\Gamma_3(\tilde{\gamma}_1+\tilde{\gamma}_2)$ & 
  $m_1 \pm \sqrt{2}m_4$ \\ \hline 
  case~ 2 & $m_2 = m_4 = 0$ & 
  $M_{R2}=m_1\bm{1}+m_3(\tilde{\gamma}_1+\tilde{\gamma}_2)$ & 
  $m_1 \pm \sqrt{2}m_3$ \\ \hline 
  case~ 3 & $m_3 = m_4 = 0$ & $M_{R3}=m_1\bm{1}+m_2\Gamma_3$ & 
  $m_1 \pm m_2$ \\ \hline 
\end{tabular}
\caption{Three cases for the mass splitting into two spinors.} 
\label{3C}
\end{center}
\end{table}

\paragraph{Case~1} ~ 

\nid
The mass matrix
\be
 M_{R1} = \left(\begin{array}{cccc} 
   m_1 & m_4 & -m_4 &   0 \\ 
   m_4 & m_1 &    0 & m_4 \\ 
  -m_4 &   0 &  m_1 & m_4 \\ 
     0 & m_4 &  m_4 & m_1\end{array}\right) 
\ee
\nid
can be diagonalized as
\be
 M_{R1}^{diag} = P_1^\dag M_{R1} P_1
 = diag(m_1-\sqrt{2}m_4, m_1-\sqrt{2}m_4, m_1+\sqrt{2}m_4, m_1+\sqrt{2}m_4),
\ee
\nid
by a unitary matrix
\be
 P_1 = \frac{1}{2\sqrt{2}} \left(\begin{array}{cccc} 
  1+i & 1-i & 1-i & 1+i \\ 
  -\sqrt{2} & -\sqrt{2} & \sqrt{2} & \sqrt{2} \\ 
  \sqrt{2}i & -\sqrt{2}i & \sqrt{2}i & -\sqrt{2}i \\ 
  1-i & 1+i & 1+i & 1-i \end{array}\right). 
\ee
\nid
Note that the matrix $P_1$ can diagonalize the rotation matrix 
$V_{12}$ simultaneously.  The unitary transformed spinor is given by  
\ba
 \Psi_{M_1}(N) &\equiv& P_1^\dag \Psi (N) = 
 \frac{1}{2\sqrt{2}} \left(\begin{array}{cccc} 
  1-i & -\sqrt{2} & -\sqrt{2}i & 1+i \\ 
  1+i & -\sqrt{2} & \sqrt{2}i & 1-i \\ 
  1+i & \sqrt{2} & -\sqrt{2}i & 1-i \\ 
  1-i & \sqrt{2} & \sqrt{2}i & 1+i \end{array}\right) 
 \left(\begin{array}{l} 
  \Psi_1 \\ \Psi_2 \\ \Psi_3 \\ \Psi_4 \end{array}\right)(N) \no 
 &=& \frac{1}{2\sqrt{2}} \left(\begin{array}{c} 
  (1-i)\Psi_1 - \sqrt{2}\Psi_2 - \sqrt{2}i\Psi_3 + (1+i)\Psi_4 \\ 
  (1+i)\Psi_1 - \sqrt{2}\Psi_2 + \sqrt{2}i\Psi_3 + (1-i)\Psi_4 \\ 
  (1+i)\Psi_1 + \sqrt{2}\Psi_2 - \sqrt{2}i\Psi_3 + (1-i)\Psi_4 \\ 
  (1-i)\Psi_1 + \sqrt{2}\Psi_2 + \sqrt{2}i\Psi_3 + (1+i)\Psi_4
 \end{array}\right)(N). 
\ea

\paragraph{Case~2} ~ 

\nid
The mass matrix
\be
 M_{R2} = \left(\begin{array}{cccc} 
   m_1 & -im_3 & -im_3 &     0 \\ 
  im_3 &   m_1 &     0 & -im_3 \\ 
  im_3 &     0 &   m_1 &  im_3 \\ 
     0 &  im_3 & -im_3 &   m_1 \end{array}\right) 
\ee
\nid
can be diagonalized as
\be
 M_{R2}^{diag} = P_2^\dag M_{R2} P_2
 = diag(m_1-\sqrt{2}m_3, m_1-\sqrt{2}m_3, m_1+\sqrt{2}m_3, m_1+\sqrt{2}m_3),
\ee
\nid
by a unitary matrix
\be
 P_2 = \frac{1}{2\sqrt{2}} 
 \pmatrix{
 -\sqrt{2}i & \sqrt{2}i & -\sqrt{2}i & -\sqrt{2}i \cr
 -1+i & 1+i & 1+i & 1-i \cr
 -1-i & 1-i & 1-i & 1+i \cr
 \sqrt{2} & \sqrt{2} & -\sqrt{2} & \sqrt{2}},
\ee
\nid
with the transformed spinor 
\be
 \Psi_{M_2}(N) \equiv P_2^\dag \Psi (N) = \frac{1}{2\sqrt{2}} 
 \left(\begin{array}{cccc} 
   \sqrt{2}i\Psi_1 - (1+i)\Psi_2 - (1-i)\Psi_3 + \sqrt{2}\Psi_4 \\ 
  -\sqrt{2}i\Psi_1 + (1-i)\Psi_2 + (1+i)\Psi_3 + \sqrt{2}\Psi_4 \\ 
   \sqrt{2}i\Psi_1 + (1-i)\Psi_2 + (1+i)\Psi_3 - \sqrt{2}\Psi_4 \\ 
   \sqrt{2}i\Psi_1 + (1+i)\Psi_2 + (1-i)\Psi_3 + \sqrt{2}\Psi_4 
 \end{array}\right)(N). 
\ee

\paragraph{Case~3} ~ 

\nid
The mass matrix 
\be
 M_{R3} = \left(\begin{array}{cccc} 
  m_1 &    0 &     0 & -im_2 \\ 
    0 &  m_1 & -im_2 &     0 \\ 
    0 & im_2 &   m_1 &     0 \\ 
 im_2 &    0 &     0 &   m_1 \end{array}\right) 
\ee
\nid
can be diagonalized as
\be
 M_{R3}^{diag} = P_3^\dag M_{R3} P_3
 = diag(m_1-m_2, m_1-m_2, m_1+m_2, m_1+m_2),
\ee
\nid
by a unitary matrix
\be
 P_3 = \frac{1}{2\sqrt{2}} \left(\begin{array}{cccc} 
  -1+i & 1-i & -\sqrt{2}i & -\sqrt{2}i \\ 
  \sqrt{2}i & \sqrt{2}i & 1-i & -1+i \\ 
  \sqrt{2} & \sqrt{2} & 1+i & -1-i \\ 
  1+i & -1-i & \sqrt{2} & \sqrt{2} \end{array}\right),
\ee
\nid
with the transformed spinor 
\be
 \Psi_{M_3}(N) \equiv P_3^\dag \Psi (N) 
  = \frac{1}{2\sqrt{2}} \left(\begin{array}{cccc} 
  -(1+i)\Psi_1 - \sqrt{2}i\Psi_2 + \sqrt{2}\Psi_3 + (1-i)\Psi_4 \\ 
  (1+i)\Psi_1 - \sqrt{2}i\Psi_2 + \sqrt{2}\Psi_3 - (1-i)\Psi_4 \\ 
  \sqrt{2}i\Psi_1 + (1+i)\Psi_2 + (1-i)\Psi_3 + \sqrt{2}\Psi_4 \\ 
  \sqrt{2}i\Psi_1 - (1+i)\Psi_2 - (1-i)\Psi_3 + \sqrt{2}\Psi_4 
 \end{array}\right)(N). 
\ee

The feature of our formulation is to keep  
the discrete rotational invariance.  After the mass splitting,  
we can find  the character of a Dirac spinor under the rotation, 
\be
 \psi(x) \rightarrow  Q\psi(R(x)) , 
\ee
\nid
where 
$Q = e^{(i\pi/4)\sigma_3}
   = \pmatrix{ e^{i\pi/4} & 0 \cr 0 & e^{-i\pi/4}}$.  
Actually in  cases 1 and 2  we can  keep 
the property of a Dirac spinor on lattice, 
\be
 \Psi(N) \rightarrow V_{12}\Psi(R(N)). 
\ee
\nid
By contrast, $\Psi(N)$ acts as a vector not as a spinor 
in case 3.  The properties  of $2$-component spinors 
under the rotation  are summarized in Table \ref{DS}.
\begin{table}
\begin{center}
\begin{tabular}{|c|c|c|}
 \hline & $P_i^\dag V_{12} P_i$ & phase factor of $V_{12}$ \\ \hline
  $\begin{array}{c} ~ \\[-5pt] {\rm case~1} \\[-5pt] ~ \end{array}$ & 
  $\left(\begin{array}{cc} Q & 0 \\ 0 & e^{i\pi}Q^\dag 
   \end{array}\right)$ & 
  $e^{i\vartheta}=e^{i\pi/2}=i$ \\ \hline 
  $\begin{array}{c} ~ \\[-5pt] {\rm case~2} \\[-5pt] ~ \end{array}$ & 
  $\left(\begin{array}{cc} Q & 0 \\ 0 & e^{i\pi}Q^\dag 
   \end{array}\right)$ & 
  $e^{i\vartheta}=e^{i\pi}=-1$ \\ \hline 
  $\begin{array}{c} ~ \\[-5pt] {\rm case~3} \\[-5pt] ~ \end{array}$ & 
  $\left(\begin{array}{cc} Q^2 & 0 \\ 0 & e^{i\pi/2}(Q^\dag)^2 
   \end{array}\right)$ & 
  $e^{i\vartheta}=e^{-i\pi/4}=(1-i)/\sqrt{2}$ 
 \\ \hline 
\end{tabular}
\caption{The properties of  Dirac spinors under the rotation.}
\label{DS}
\end{center}
\end{table}

\section{Pole Analysis and 2-point Functions}    
Our adding terms do not commute with the staggered Dirac operator.  
As a result,  our analysis in the  previous section 
is insufficient to split masses.  We must proceed in 
the pole analysis of  the theory because a pole mass is physical.  
For the help,  Dirac fields are  Fourier transformed as 
\be
 \Psi_r(N)=\displaystyle\int_{-\pi}^{+\pi}\frac{d^2p}{(2\pi)^2}\tilde{\Psi}_r(p)e^{ipN}, \quad 
 \bar{\Psi}_r(N)=\displaystyle\int_{-\pi}^{+\pi}\frac{d^2p}{(2\pi)^2}\tilde{\bar{\Psi}}_r(p)e^{ipN}, 
\ee
\nid
and the action becomes 
\ba
 S_{st}&=& \sum_{N,N^\prime,r,r^\prime,\mu,\vec{\tau}} 
  \bar{\Psi}_r(N)(D_{\mu}^{\vec{\tau}})_{(N,N^\prime)} 
  (\Gamma_{\mu,\vec{\tau}})_{(r,r^\prime)}\Psi_{r^\prime}(N^\prime) 
 \no &=& \sum_{r,r^\prime}\int_{-\pi}^{+\pi}\frac{d^2p}{(2\pi)^2} 
  \tilde{\bar{\Psi}}_r(-p)\left[\sum_\mu\left\{i\gamma_\mu\sin p_\mu 
  +i\tilde{\gamma}_\mu(1-\cos p_\mu)\right\}\right]_{(r,r^\prime)} 
  \tilde{\Psi}_{r^\prime}(p). 
\ea
\nid
The staggered Dirac operator is explicitly written as 
\ba
 D_{st}(p)&=&\sum_\mu\left\{i\gamma_\mu\sin 
 p_\mu+i\tilde{\gamma}_\mu(1-\cos p_\mu)\right\} 
 \no &=& \left(\begin{array}{cccc} 
  0 & is_2 + c_2 & is_1 + c_1 & 0 \\ 
  is_2 - c_2 & 0 & 0 & is_1 + c_1 \\ 
  is_1 - c_1 & 0 & 0 & -is_2 - c_2 \\ 
  0 & is_1 - c_1 & -is_2 + c_2 & 0 \end{array}\right), 
\ea
\nid
where $s_i \equiv \sin p_i$ and $c_i \equiv 1-\cos p_i$ for  
$i=1, 2$, respectively. 

Our steps  to find a pole mass are as follows:
(i) set $p_1=0$ and $p_2= i \kappa~({\rm pure ~imaginary})$ of 
the inverse propagator  $D^{-1}$ in the momentum representation 
where our rotationally invariant operators~(\ref{4rot-inv-ope}) are 
included; (ii)  calculate  four eigenvalues $\lambda$ of 
$D^{-1}$; (iii) find values of $\kappa$ in setting $\lambda=0$.  Four 
values of $\kappa$  equal to pole masses.  As mentioned in 
sections 2 and 3, we keep the rotational invariance in our action and 
generate two Dirac spinors with different masses.  The light Dirac 
mass is denoted as $m_l$ and the heavy one is done as  $m_h$.  
We  note that  parameters 
$m_1$, $m_2^{\prime}\equiv -i m_2$, $m_3^{\prime}\equiv -i m_3$ and 
$m_4$ are real  to obtain real pole masses.
\paragraph{Case~1} ~ 

\nid
The improved staggered Dirac operator  is given by 
\ba
 D_1^{imp} &\equiv& D_{st}(p) + M_{R1} \no &=& \sum_\mu
 \left\{i\gamma_\mu\sin p_\mu+i\tilde{\gamma}_\mu(1-\cos p_\mu)\right\} 
  + m_1 + m_4 \Gamma_3(\tilde{\gamma}_1 + \tilde{\gamma}_2) 
 \no &=& \left(\begin{array}{cccc} 
  m_1 & is_2 + c_2 +m_4 & is_1 + c_1 -m_4 & 0 \\ 
  is_2 - c_2 +m_4 & m_1 & 0 & is_1 + c_1 +m_4 \\ 
  is_1 - c_1 -m_4 & 0 & m_1 & -is_2 - c_2 +m_4 \\ 
  0 & is_1 - c_1 +m_4 & -is_2 + c_2 +m_4 & m_1 \end{array}\right). 
\label{DImp1}
\ea
\nid
Setting $p_1=0$, we obtain  eigenvalues of Eq.~(\ref{DImp1}) as 
\ba
 \lambda &=& m_1 \pm 
  \sqrt{2m_4^2-4\sin^2 \displaystyle\frac{p_2}{2} \pm 
  \sqrt{-4m_4^2\left(\sin^2 p_2+4\sin^2 \displaystyle\frac{p_2}{2} 
 \right)}}. 
\label{lam1}
\ea
\nid
Taking $p_2=i\kappa$ at $\lambda=0$ in Eq.~(\ref{lam1}), we have 
the equation for the pole mass
\be
 16(1-m_4^2)\sinh^4 \displaystyle\frac{\kappa}{2} 
 - 8(m_1^2+2m_4^2)\sinh^2 \displaystyle\frac{\kappa}{2} 
 + (m_1^2-2m_4^2)^2=0. 
\label{eqn:limit}
\ee
\nid
Hence we find pole masses for $|m_4| < 1$ as 
\be
 \kappa=\left\{\begin{array}{ll}
 \pm\ln\left[\displaystyle\frac{2+m_1^2-\sqrt{A}}{2(1-m_4^2)} +
 \sqrt{\left\{\frac{2+m_1^2-\sqrt{A}}{2(1-m_4^2)}\right\}^2 - 1} \
 \right] \equiv \pm m_l , 
\\[9mm]
 \pm\ln\left[\displaystyle\frac{2+m_1^2+\sqrt{A}}{2(1-m_4^2)} +
 \sqrt{\left\{\frac{2+m_1^2+\sqrt{A}}{2(1-m_4^2)}\right\}^2 - 1} \
 \right] \equiv \pm m_h , 
\end{array}\right. 
\label{pole-c1}
\ee
where $A \equiv m_4^2\left\{8m_1^2+(m_1^2-2m_4^2)^2 \right\}$.  

The mass is still splitting in the pole analysis for 
the improved staggered action.  {}From Eq.~(\ref{pole-c1}), it is 
found that  we can take a limit $|m_h| \rightarrow \infty$ for 
arbitrary $m_l$ by performing $\epsilon \rightarrow 0$ in 
an expression $m_4^2=1-\epsilon\;(0<\epsilon\ll1)$.  

\paragraph{Case~2} ~ 

\nid
The improved staggered Dirac operator is given by 
\ba
 D_2^{imp} &\equiv& D_{st}(p) + M_{R2}^\prime \no 
 &=& \sum_\mu\left\{i\gamma_\mu\sin 
  p_\mu+i\tilde{\gamma}_\mu(1-\cos p_\mu)\right\}+m_1 
  + im_3^\prime (\tilde{\gamma}_1 + \tilde{\gamma}_2) \no 
 &=& \sum_\mu\left\{i\gamma_\mu\sin 
  p_\mu+i\tilde{\gamma}_\mu(1+m_3^\prime-\cos p_\mu)\right\}+m_1 \no 
 &=& \left(\begin{array}{cccc} 
  m_1 & is_2 + c_2^\prime  & is_1 + c_1^\prime  & 0 \\ 
  is_2 - c_2^\prime  & m_1 & 0 & is_1 + c_1^\prime  \\ 
  is_1 - c_1^\prime  & 0 & m_1 & -is_2 - c_2^\prime \\ 
  0 & is_1 - c_1^\prime  & -is_2 + c_2^\prime  & m_1 
 \end{array}\right), 
\label{DImp2}
\ea
\nid
where $c_i^\prime \equiv 1+m_3^\prime-\cos p_i$.  Eigenvalues of 
Eq.~(\ref{DImp2}) with $p_1=0$ are 
\ba
 \lambda&=&m_1\pm\sqrt{-4(1+m_3^\prime)\sin^2
  \displaystyle\frac{p_2}{2} - 2(m_3^\prime)^2}. 
\ea
\nid
Setting $p_2 = i\kappa$ and $\lambda=0$,  the pole mass 
is satisfied with 
\be
 \sinh^2 \displaystyle\frac{\kappa}{2} = 
 \frac{m_1^2+2(m_3^\prime)^2}{4(1+m_3^\prime)}. 
\ee
\nid
Solutions of this equation under  $-1<m_3^\prime$ are 
\be
 \kappa = \pm2\ln\left[\displaystyle 
  \sqrt{\frac{m_1^2+2(m_3^\prime)^2}{4(1+m_3^\prime)}} + 
  \sqrt{\frac{m_1^2+2(m_3^\prime)^2}{4(1+m_3^\prime)}+1} \ 
 \right].
\label{lam2}
\ee
\nid
The pole mass remains degenerate because the improved term 
$m_3^\prime (\tilde{\gamma}_1 + \tilde{\gamma}_2)$ is absorbed into 
the kinetic term. 

\paragraph{Case~3} ~ 

\nid
The improved staggered Dirac operator is given by 
\ba
 D_3^{imp} &\equiv& D_{st}(p) + M_{R3}^\prime \no 
 &=& \sum_\mu\left\{i\gamma_\mu\sin 
  p_\mu+i\tilde{\gamma}_\mu(1-\cos p_\mu)\right\}+m_1 
  + im_2^\prime \Gamma_3 \no 
 &=& \left(\begin{array}{cccc} 
  m_1 & is_2 + c_2  & is_1 + c_1  & m_2^\prime \\ 
  is_2 - c_2  & m_1 & m_2^\prime & is_1 + c_1 \\ 
  is_1 - c_1  & -m_2^\prime & m_1 & -is_2 - c_2 \\ 
  -m_2^\prime & is_1 - c_1  & -is_2 + c_2  & m_1 
 \end{array}\right). 
\label{DImp3}
\ea
\nid
Eigenvalues of Eq.~(\ref{DImp3}) with $p_1=0$ are 
\ba
 \lambda&=&m_1\pm\sqrt{-4\sin^2 \displaystyle\frac{p_2}{2} - 
  (m_2^\prime)^2 \pm 4m_2^\prime \sin^2 \displaystyle\frac{p_2}{2}}. 
\ea
\nid
Setting $p_2 = i\kappa$ and $\lambda=0$, the pole mass 
is satisfied with 
\be
 \sinh^2 \displaystyle\frac{\kappa}{2} = 
 \frac{m_1^2+(m_2^\prime)^2}{4(1\pm m_2^\prime)}. 
\ee
Solutions of this equation under  $-1<m_2^\prime<1$  are 
\be
 \kappa=\left\{\begin{array}{ll}
 \pm 2\ln \left[
  \sqrt{\displaystyle\frac{m_1^2+(m_2^\prime)^2}{4(1+
  m_2^\prime)}}+\sqrt{\displaystyle\frac{m_1^2+(m_2^\prime)^2}
  {4(1+m_2^\prime)}+1} \ \right] , \\[6mm] 
 \pm 2\ln \left[
  \sqrt{\displaystyle\frac{m_1^2+(m_2^\prime)^2}{4(1-
  m_2^\prime)}}+\sqrt{\displaystyle\frac{m_1^2+(m_2^\prime)^2}
  {4(1-m_2^\prime)}+1} \ \right] . 
 \end{array}\right. 
\label{lam3}
\ee
This  case  allows  pole masses  to split 
although the rotational property of the eigenmode 
is not a spinor from the discussion of the previous section.  

In summary, we find that the improved staggered Dirac operator 
for cases 1 and 3 can split the degenerate mass through 
the analysis of the inverse propagator.  However, that of case 2 
cannot do because the additional effect is absorbed into 
the kinetic term of the original staggered Dirac operator.  
In addition, we should mention that there is no massless mode 
in cases 2 and 3 from Eqs.~(\ref{lam2}) and~(\ref{lam3}).  These 
nontrivial results are originated from the fact that 
the rotationally invariant mass terms do not commute with 
the staggered Dirac operator.  Finally note that it is possible 
to take the light mass $m_l$ to zero  by tuning $m_1$ and $m_4$ 
only in case 1.  Focusing on this particular situation, we 
discuss about massless and infinity modes in the next section.

\section{Massless and Infinity Modes of Rotationally Invariant Action 
for Viable Even-Odd Separation} 
In our two-dimensional lattice formulation, a matrix 
\be
 \Gamma_5 = \gamma_1\gamma_2\tilde{\gamma}_1\tilde{\gamma}_2 = 
 \left(\begin{array}{cccc} 
  1&0&0&0 \\ 0&-1&0&0 \\ 0&0&-1&0 \\ 0&0&0&1 
 \end{array}\right) 
\ee
\nid
can define a  chiral projection for  a Dirac spinor because of  
$\{D_{st}, \Gamma_5 \} = 0$.  {}From the explicit 
matrix representation, the positive chiral mode is put on even sites 
and  the negative mode corresponds to odd sites, 
\be
 \Psi_e(N) \equiv \frac{1+\Gamma_5}{2}\Psi (N) = 
 \left(\begin{array}{c} 
  \Psi_1 \\ 0 \\ 0 \\  \Psi_4 \end{array}\right)(N), \quad 
 \Psi_o(N) \equiv \frac{1-\Gamma_5}{2}\Psi (N) = 
 \left(\begin{array}{c} 
  0 \\ \Psi_2  \\ \Psi_3 \\ 0\end{array}\right)(N). 
\ee
\nid
We note that the chiral projection is not discrete 
rotationally invariant since  $[V_{12},  \Gamma_5] \ne 0$.  

This (even-odd) chiral property of staggered fermions holds 
in general $D$-dimensional cases.  First of all, we define 
chiral projection operators 
\be
 P_L \equiv \frac{1-\Gamma_{2D+1}}{2}, \quad 
 P_R \equiv \frac{1+\Gamma_{2D+1}}{2}, 
\ee
\nid
and a Dirac spinor  $\Psi$ is projected out as 
\be
 \Psi_L \equiv  P_L \Psi, \quad  \Psi_R \equiv  P_R \Psi. 
\ee
\nid
The kinetic term of  a staggered Dirac fermions lagrangian  
can be written as 
\be
 {\mathcal L}_{st} = 
 \bar{\Psi}_L D_{st} \Psi_L + \bar{\Psi}_R D_{st}\Psi_R. 
\ee
\nid
Among operators discussed in sections 3 and 4, ${\mathcal M}_3$ and 
${\mathcal M}_4$ are chirally invariant  but ${\mathcal M}_1$ and 
${\mathcal M}_2$ change  chiralities of Dirac spinors.  Therefore, 
${\mathcal M}_1$ and ${\mathcal M}_2$ can  construct Dirac mass terms 
while  ${\mathcal M}_3$ and ${\mathcal M}_4$ can do 
Majorana mass terms.  Although we can define the formal discussion of 
the chirality just as normal chirality $\gamma_5$, this definition of 
the chirality depends on  a special lattice frame 
because $[ V_{\mu\nu}, \Gamma_{2D+1}  ]\ne 0$ where 
$V_{\mu\nu}$ means the $\mu\nu$-rotation by $\pi/2$.  

Without mention about  the chirality, we can define  
a massless mode and can throw up the mass of a heavy mode  
to infinity in case 1 of section 4.  
Solutions of Eq.~(\ref{eqn:limit}) 
under the massless condition $m_1^2=2m_4^2$ are 
determined as 
\be
 \sinh^2 \frac{m_l}{2}=0 , 
\label{massless-condition} 
\ee
\be
 \sinh^2 \displaystyle\frac{m_h}{2} = \frac{2m_4^2}{1-m_4^2} .  
\label{inf-condition}
\ee

It must be noted that the eigenmode of the Dirac operator 
around a massless pole  is not orthogonal to that 
around a heavy mass pole because their Dirac operators are 
different from each other.  The massless modes 
for $m_4 >0$ are explicitly written by 
\be
 \left( \begin{array}{c} 
  1 + \sqrt{2} \\ -1 - \sqrt{2} \\ 1 \\ 1 \end{array} \right) 
 \quad , \quad  \left( \begin{array}{c} 
  1 - \sqrt{2} \\ 1 - \sqrt{2} \\ -1 \\ 1 \end{array} \right) . 
\ee
\nid
In $m_4 < 0$ case,  they are  expressed as  
\be
 \left( \begin{array}{c} 
  1 - \sqrt{2} \\ -1 + \sqrt{2} \\ 1 \\ 1 \end{array} \right) 
 \quad , \quad  \left( \begin{array}{c} 
  1 + \sqrt{2} \\ 1 + \sqrt{2} \\ -1 \\ 1 \end{array} \right) . 
\ee
\nid
In order to get the eigenmode for the heavy mass, we take 
$p_1=0$ and  $p_2=im_h$ for Eq.~(\ref{DImp1}).  
$D_1^{imp} (p_1=0,\ p_2=im_h)$ is given by a form 
\ba
 \left(\begin{array}{cccc} 
  m_1 & 1 + m_4 - e^{m_h} & -m_4 & 0 \\ 
  -1 + m_4 + e^{-m_h} & m_1 & 0 & m_4 \\ 
  -m_4 & 0 & m_1 & -1 + m_4 + e^{m_h} \\ 
  0 & m_4 & 1 + m_4 - e^{-m_h} & m_1 \end{array}\right). 
\label{DImp1-mh-p}
\ea
\nid
Furthermore, from Eq.~(\ref{inf-condition}), we find that 
eigenmodes corresponding to the heavy mass are given by 
\be
 \left( \begin{array}{c} 
  \displaystyle\frac{1+m_4}{m_4} 
   \left\{m_4(1+m_4) + \sqrt{2m_4^2(1+m_4^2)}\right\} \\[3mm] 
  \displaystyle\frac{1-m_4}{m_4} 
   \left(\sqrt{2m_4^2} + m_4\sqrt{1+m_4^2}\right) \\[3mm] 
  -(1-m_4) \left(\sqrt{2m_4^2} + \sqrt{1+m_4^2}\right) \\[3mm] 
  1-m_4^2 \end{array} \right), 
\label{em-p}
\ee
\nid
for $m_h > 0$ and 
\be
 \left( \begin{array}{c} 
  1-m_4^2 \\[3mm] 
  -(1-m_4) \left(\sqrt{2m_4^2} + \sqrt{1+m_4^2}\right) \\[3mm] 
  -\displaystyle\frac{1-m_4}{m_4} 
   \left(\sqrt{2m_4^2} + m_4\sqrt{1+m_4^2}\right) \\[3mm] 
  -\displaystyle\frac{1+m_4}{m_4} 
   \left\{m_4(1+m_4) + \sqrt{2m_4^2(1+m_4^2)}\right\} 
   \end{array} \right), 
\label{em-m}
\ee
\nid
for $m_h < 0$.  

To decouple the heavy mode, we can throw the mass  
up to  infinity.  Actually from Eqs.~(\ref{massless-condition}) and  
(\ref{inf-condition}), we  can realize massless and infinity modes as 
Table~\ref{limitmode}  simultaneously.  {}From these results, 
it is found that 
infinity modes can be  separately  put  on even or odd sites.
\begin{table}[h]
\begin{center}
\begin{tabular}{|c|c|c|} 
 \hline & massless modes & infinity modes \\ \hline 
   $\begin{array}{c} ~ \\[4mm] m_4 > 0 \\[4mm] ~  \end{array}$ 
 & $\left(\begin{array}{c} 
     1+\sqrt{2} \\ -1-\sqrt{2} \\ 1 \\ 1 
    \end{array}\right) , \quad 
    \left(\begin{array}{c} 
     1-\sqrt{2} \\ 1-\sqrt{2} \\ -1 \\ 1 
    \end{array}\right)$ 
 & $\left(\begin{array}{c} 1 \\ 0 \\ 0 \\ 0 \end{array}\right) , \quad 
    \left(\begin{array}{c} 0 \\ 0 \\ 0 \\ 1 \end{array}\right)$ 
 \\ \hline 
   $\begin{array}{c} ~ \\[4mm] m_4 < 0 \\[4mm] ~  \end{array}$ 
 & $\left(\begin{array}{c} 
     1-\sqrt{2} \\ -1+\sqrt{2} \\ 1 \\ 1 
    \end{array}\right) , \quad 
    \left(\begin{array}{c} 
     1+\sqrt{2} \\ 1+\sqrt{2} \\ -1 \\ 1 
    \end{array}\right)$
 & $\left(\begin{array}{c} 0 \\ 0 \\ 1 \\ 0 \end{array}\right) , \quad 
    \left(\begin{array}{c} 0 \\ 1 \\ 0 \\ 0 \end{array}\right) $ 
 \\ \hline 
\end{tabular}
\caption{Eigenvectors of the improved Dirac operator in case 1 with  
$m_1^2 = 2 m_4^2$.  Expressions of infinity modes are given by 
substituting $m_4^2 = 1$ for Eqs.~(\ref{em-p}) and (\ref{em-m}).}
\label{limitmode}
\end{center}
\end{table}

\section{Summary and Discussion}
We have studied  the mass splitting of two-dimensional 
staggered fermions based on the $SO(4)$ Clifford algebra.  Introducing 
four rotationally invariant operators,
we have analyzed three types of improved staggered Dirac operators 
and found one possibility (case 1) for taking a single mode 
in a two-dimensional free theory.  The case keeps the splitting 
not only in the analysis of the mass matrix itself 
but also in the pole analysis including the kinetic term.  According
to the improvement with respect to the rotational invariance, the  
derived $2$-component modes can be regarded as the ordinary spinor  
under the rotation by $\pi/2$.  Furthermore, one can find 
a massless mode in the case  unexpectedly.  The formal  $\Gamma_5$ 
chiral projection which means even-site and odd-site separation of  
fermion modes  is not consistent with the rotational invariance of 
a staggered Dirac action.  Nevertheless,  massless and infinity 
mode-representations in the case  realize 
even-odd separation of the infinity mode.  

Our future tasks are  analyses of interacting theories  and 
the extension of our approach to  four dimensions.  In particular, 
it is crucial that 
the stability for the massless condition under quantum corrections 
by gauge interactions.  Namely, in the case that the theory 
is not stable, it may be uninteresting that one needs a fine-tuning 
of the additional mass parameter as in  Wilson fermions.  
For the infinity mode, it is very interesting 
if the even-odd separation is induced  in  staggered fermions 
even when we consider  interaction effects. 

\vspace{1cm}

Acknowledgments 

The authors thank Hiroshi Suzuki for valuable comments at  
``Niigata-Yamagata School'' (YITP-S-05-02).  H. So  is stimulated 
for  staggered fermions during the YITP workshop on 
``Actions and Symmetries in Lattice Gauge Theory" (YITP-W-05-25).  
He also would like to thank  David B. Kaplan 
for discussions about staggered fermions  at that workshop, 
which partially motivated the present work.  
This work is supported in part by the Grants-in-Aid for Scientific
Research No.  17540242 from the Japan Society for the
Promotion of Science.

\clearpage

\end{document}